\documentclass[twocolumn,amsmath,amssymb,a4paper,prb,superscriptaddress,floatfix]{revtex4-1}
\usepackage[dvipdfmx]{graphicx}
\usepackage{natbib}
\usepackage{multirow}
\usepackage{amsmath}
\usepackage{bm}
\usepackage{mathrsfs}
\usepackage{url}
\usepackage{color}
\usepackage{ulem}
\begin{document}

\title{Lattice thermal conductivities of two SiO$_2$ polymorphs by
first-principles calculation and phonon Boltzmann transport equation}

\author{Keiyu Mizokami}
\affiliation{Department of Materials Science and
Engineering, Kyoto University, Sakyo, Kyoto 606-8501, Japan}

\author{Atsushi Togo}
\email{togo.atsushi@gmail.com}
\affiliation{Center for Elements Strategy Initiative for Structural
Materials, Kyoto University, Sakyo, Kyoto 606-8501, Japan}

\author{Isao Tanaka}
\affiliation{Department of Materials Science and
Engineering, Kyoto University, Sakyo, Kyoto 606-8501, Japan}
\affiliation{Center for Elements Strategy Initiative for Structural
Materials, Kyoto University, Sakyo, Kyoto 606-8501, Japan}
\affiliation{Nanostructures Research Laboratory, Japan Fine Ceramics
Center, Atsuta, Nagoya 456-8587, Japan}
\affiliation{Center for Materials Research by Information Integration,
National Institute for Materials Science, Tsukuba 305-0047, Japan}

\begin{abstract}
 Lattice thermal conductivities of two SiO$_2$ polymorphs, i.e.,
 $\alpha$-quartz (low) and $\alpha$-cristobalite (low), were studied
 using first-principles anharmonic phonon calculation and linearized
 phonon Boltzmann transport equation. Although $\alpha$-quartz and
 $\alpha$-cristobalite have similar phonon densities of states, phonon
 frequency dependencies of phonon group velocities and lifetimes are
 dissimilar, which results in largely different anisotropies of the
 lattice thermal conductivities. For $\alpha$-quartz and
 $\alpha$-cristobalite, distributions of the phonon lifetimes effective
 to determine the lattice thermal conductivities are well described by
 energy and momentum conservations of three phonon scatterings weighted
 by phonon occupation numbers and one parameter that represents the
 phonon-phonon interaction strengths.
\end{abstract}

\maketitle

\section{Introduction}
Recent computing power has enabled quantitative and systematic
calculation of lattice thermal conductivity by using the combination of
first-principles calculations and solutions of linearized phonon
Boltzmann transport equation.\cite{Peierls-1929, Peierls-1935,
Hardy-1970, Omini-1997, Deinzer-2003, Broido-kappa-2005,
Broido-kappa-2007, Ward-kappa-2009, Turney-2009, Chernatynskiy-2010,
Laurent-LBTE-2013, Fugallo-2013, Hellman-2014, ShengBTE-2014, phono3py,
Tadano-2015, Cepellotti-2016, Lindsay-2016, Fugallo-2018} In this study,
we applied this calculation to $\alpha$-quartz (low) and
$\alpha$-cristobalite (low) of SiO$_2$.

\begin{figure}[ht]
  \begin{center}
    \includegraphics[width=1.00\linewidth]{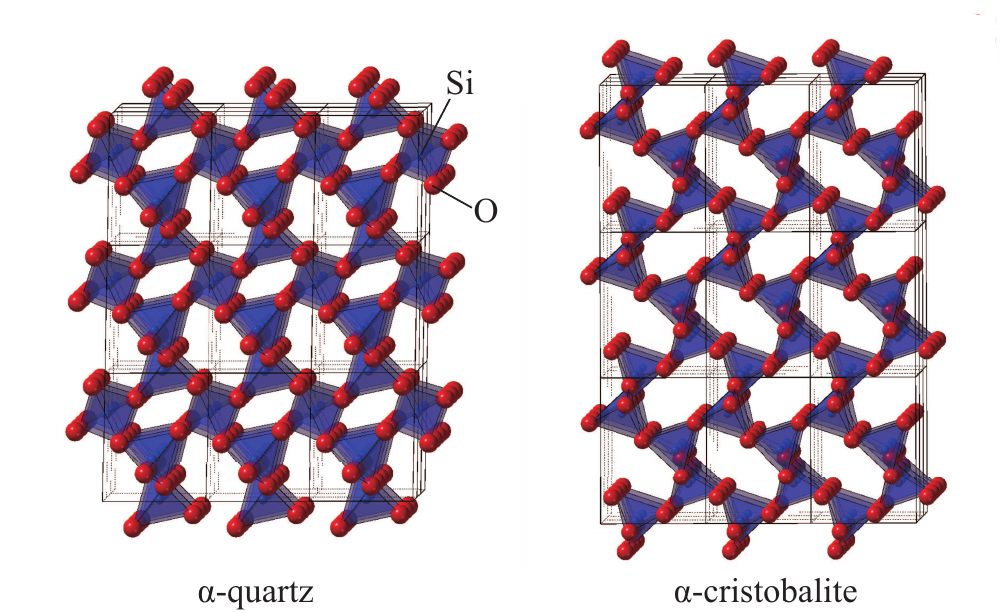} \caption{(color
    online) Crystal structures of $\alpha$-quartz (left) and
    $\alpha$-cristobalite (right). The space-group types are $P3_221$
    and $P4_12_12$,
    respectively. \label{fig:crystal-structures}}
  \end{center}
\end{figure}

SiO$_2$ exhibits many polymorphs including $\alpha$-quartz and
$\alpha$-cristobalite whose crystal structures are shown in
Fig.~\ref{fig:crystal-structures}. The numbers of atoms in the unit
cells ($n_\text{a}$) are 9 and 12, respectively. Both are made of
SiO$_4$ tetrahedra connected by their vertices. Si atom is located at
the center of each tetrahedron and O atoms are at the vertices. The
difference of these crystal structures is described by the patterns of
the tetrahedron linkages. SiO$_4$ tetrahedra are more densely packed in
$\alpha$-quartz.  As a result, the volume per formula unit is more than
ten percent smaller in $\alpha$-quartz. Their lattice
parameters~\cite{Antao-2008-quartz-explat,Pluth-1985-cristobalite-explat}
are shown in Table.~\ref{table:latparams-wrt-settings}.  Their
space-group types are $P3_221$ (trigonal) for $\alpha$-quartz and
$P4_12_12$ (tetragonal) for $\alpha$-cristobalite.  Both of them in
principle have anisotropic thermal conductivity tensors with two
independent elements, $\kappa_{xx}$ and $\kappa_{zz}$. Although
$\kappa_{xx}$ and $\kappa_{zz}$ of $\alpha$-quartz were
reported,\cite{Kanamori-1968} only its average value is known for
$\alpha$-cristobalite.\cite{Kunugi-1972}


The aim of this study is to understand their difference in lattice
thermal conductivity. Indeed $\alpha$-quartz shows much larger
anisotropy in lattice thermal conductivity than $\alpha$-cristobalite as
presented in this study. This was investigated from microscopic
properties due to phonons. By the long range interaction among atoms and
softer low frequency phonon modes, we were required to conduct more
careful calculations than that usually we do in conjunction with our
software development.\cite{phonopy-project, phono3py-project} These
computational details and comparisons of calculations with experiments
are presented in Sec.~\ref{sec:method-of-calculation}. Results of lattice
thermal conductivity calculations and their analysis are
presented in Sec.~\ref{sec:results}. We show similarity and
dissimilarity between $\alpha$-quartz and $\alpha$-cristobalite in
densities of lattice thermal conductivities and distributions of phonon
properties as a function of phonon frequency. Then the characteristics of
three phonon scatterings are discussed.

\section{Method of calculation}
\label{sec:method-of-calculation}

\subsection{Computational details}
\label{sec:computational-details}

We solved linearized phonon Boltzmann transport equation with
single-mode relaxation time approximation.\cite{Physics-of-phonons,
phono3py} We abbreviate this approximation as RTA. For the phonon and
lattice thermal conductivity calculations, we employed
\texttt{phonopy}~\cite{phonopy} and \texttt{phono3py}~\cite{phono3py}
software packages. Unless specially denoted, $q$-point sampling meshes
of $19\times 19\times 19$ and $19\times 19\times 14$ were used for the
lattice thermal conductivity calculations of $\alpha$-quartz and
$\alpha$-cristobalite, respectively. The isotope scattering effect
calculated by the second-order perturbation
theory~\cite{Tamura-isotope-lifetime-1983,phono3py} was found negligibly
small. Therefore it was not included.

The experimental lattice parameters of
$\alpha$-quartz\cite{Antao-2008-quartz-explat} and
$\alpha$-cristobalite\cite{Pluth-1985-cristobalite-explat} were used for
all calculations in Sec.~\ref{sec:results}. Choice of the lattice
parameters can have a large impact to the lattice thermal conductivity
since it is known that decreasing (increasing) lattice parameters
increases (decreases) lattice thermal conductivity as has been well
studied as pressure dependence of lattice thermal conductivity for many
years.\cite{Bridgman-1924, Alm-1974, Slack-1979, Gerlich-1982,
Ross-1984} In Sec.~\ref{sec:convergence}, we present calculated lattice
thermal conductivity values obtained using experimental and calculated
lattice parameters.

Second- and third-order force constants were calculated using the
supercell approach with finite atomic displacements of 0.03
\AA.\cite{Laurent-phph-2011, phono3py} The supercells of $6\times
6\times 6$ (1944 atoms) and $4\times 4\times 4$ (768 atoms) of the unit
cells were used for the calculations of the second-order force constants
of $\alpha$-quartz and $\alpha$-cristobalite, respectively. Use of
larger supercells is in general important to compute phonon-phonon
scattering channels with better accuracy. For $\alpha$-quartz, it
was necessary to take into account the long-range interaction to remove
imaginary acoustic modes in the vicinity of $\Gamma$-point. We expect
real-space interaction range among three atoms effective for lattice
thermal conductivity is relatively shorter than that of the second-order
force constants. Therefore, for the third-order force constants, we
chose $2\times 2\times 2$ supercells (72 and 96 atoms). Our supercell
choices for $\alpha$-quartz and $\alpha$-cristobalite are considered
reasonable after the examinations as presented in
Sec.~\ref{sec:convergence}.

Running many supercell first-principles calculations for the third-order
force constants is the most computationally demanding part throughout
the lattice thermal conductivity calculation. To omit the computations
of parts of force constants in some means, e.g., using real-space cutoff
distance, can ease its total computational demand. However we filled all
elements of the supercell force constants. Nevertheless our attempts and
remarks on using the cutoff distance for computing third-order force
constants, that we avoided, are presented in Appendix.

Non-analytical term correction~\cite{Pick-1970,Gonze-1994,Gonze-1997}
was applied to dynamical matrices to treat long range dipole-dipole
interactions. Though impact of non-analytical term correction to lattice
thermal conductivity is often negligible for crystals containing number
of atoms in their unit cells such as $\alpha$-quartz (9 atoms) and
$\alpha$-cristobalite (12 atoms), it turned out to be useful for
$\alpha$-quartz to remove imaginary acoustic modes near $\Gamma$-point
in conjunction with using the larger supercell.

For the first-principles calculations, we employed the plane-wave basis
projector augmented wave method~\cite{PAW-Blochl-1994} within the
framework of density functional theory (DFT) as implemented in the VASP
code.\cite{VASP-Kresse-1995,VASP-Kresse-1996,VASP-Kresse-1999} The
generalized gradient approximation of Perdew, Burke, and Ernzerhof
revised for solids (PBEsol)~\cite{PBEsol} was used as the exchange
correlation potential. A plane-wave energy cutoff of 520 eV was
employed. The radial cutoffs of the PAW datasets of Si and O were 1.90
and 1.52 \AA, respectively. The 3s and 3p electrons for Si and the 2s
and 2p electrons for O were treated as valence and the remaining
electrons were kept frozen.
Reciprocal spaces of the $\alpha$-quartz supercells used for the
calculations of the third- and second-order force constants were sampled
by the $3\times 3\times 3$ mesh and at only $\Gamma$-point,
respectively. The former mesh was shifted by a half grid distance in $c^*$
direction from the $\Gamma$-point centered mesh. For the
$\alpha$-cristobalite supercells, the reciprocal spaces were sampled by
the $2\times 2\times 2$ and $1\times 1\times 1$ meshes with half grid
shifts along all three directions from the $\Gamma$-point centered
meshes, respectively. To obtain atomic forces, the total energies were
minimized until the energy convergences became less than $10^{-8}$ eV.

Static dielectric constant tensors and Born effective charge tensors
were calculated from density functional perturbation theory as
implemented in the VASP code.\cite{Gajdos-2006,Wu-2005} These tensors
were symmetrized by their space-group and crystallographic-point-group
operations. A sum rule was applied to the Born effective charge tensors
following Ref.~\onlinecite{Gonze-1997}. For these calculations, the
plane-wave cutoff energy of 600 eV was used. The reciprocal spaces of
the $\alpha$-quartz and $\alpha$-cristobalite were sampled by the
$12\times 12\times 12$ and $8\times 8\times 8$ $k$-point sampling
meshes, respectively. The former mesh was shifted by a half grid
distance along $c^*$ direction and the later mesh was shifted by half
grid distances along all three directions from the $\Gamma$-point
centered meshes.

\subsection{Choices of exchange correlation potentials and convergence criteria}
\label{sec:convergence}

We performed series of lattice thermal conductivity calculations against
different exchange correlation potentials, solutions of linearized
phonon Boltzmann transport equation, and convergence criteria. We
present our calculation results on them. After these examinations, we
chose the calculation settings described in
Sec.~\ref{sec:computational-details}, which are considered to give
results accurate enough for our discussion.

\begin{table}[ht]
 \caption{\label{table:latparams-wrt-settings} Experimental and
 calculated lattice parameters of $\alpha$-quartz and
 $\alpha$-cristobalite. For the calculations, PBEsol and LDA exchange
 correlation potentials were used and compared.}
 \begin{ruledtabular}
  \begin{tabular}{llcc}
    & & a (\AA) & c (\AA) \\
   \hline
   \multirow{3}{*}{$\alpha$-quartz} & Exp.\footnotemark[1] & 4.913 & 5.405 \\
    & Calc./PBEsol & 4.960 & 5.453 \\
    & Calc./LDA & 4.873 & 5.374 \\
   \hline
   \multirow{3}{*}{$\alpha$-cristobalite} & Exp.\footnotemark[2] & 4.971 & 6.928 \\
    & Calc./PBEsol & 5.045 & 7.036 \\
    & Calc./LDA & 4.956 & 6.887
  \end{tabular}
   \footnotetext[1]{Ref.~\onlinecite{Antao-2008-quartz-explat}.}
   \footnotetext[2]{Ref.~\onlinecite{Pluth-1985-cristobalite-explat}.}
 \end{ruledtabular}
\end{table}

In Table~\ref{table:latparams-wrt-settings}, the experimental lattice
parameters~\cite{Antao-2008-quartz-explat,Pluth-1985-cristobalite-explat}
of $\alpha$-quartz and $\alpha$-cristobalite and those optimized by
calculations are presented. For the calculations, we employed the
exchange correlation potentials of PBEsol and local density
approximation (LDA).\cite{LDA} Thermal expansion was not considered in
the calculations. Although the calculations show good agreements with the
experimental values, we can find that those with PBEsol and LDA slightly
overestimate and underestimate the experimental values, respectively.

\begin{figure}[ht]
  \begin{center}
    \includegraphics[width=0.90\linewidth]{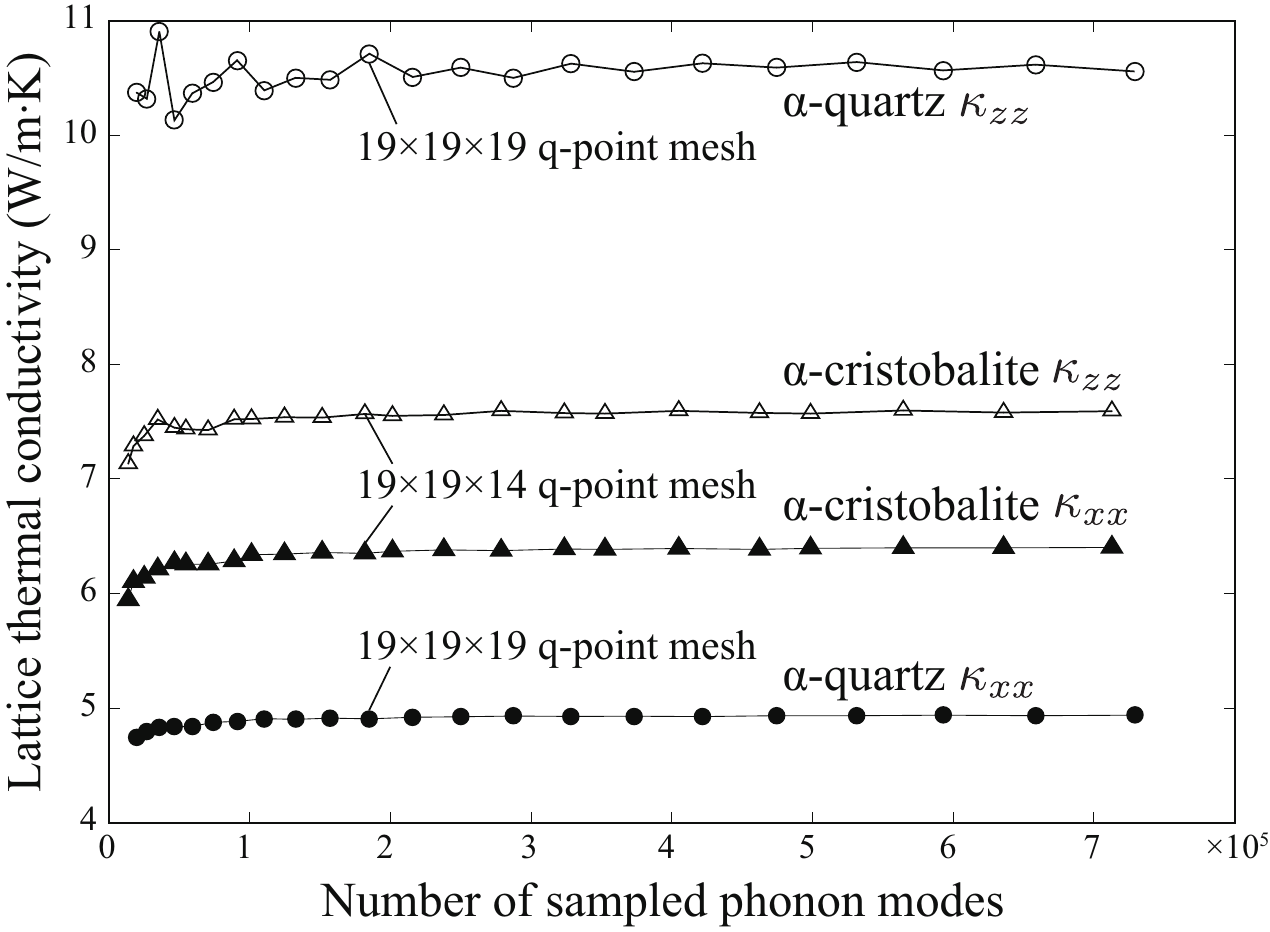} \caption{
   Lattice thermal conductivities of $\alpha$-quartz and
   $\alpha$-cristobalite calculated at 300 K with different $q$-point
   sampling meshes using the PBEsol exchange correlation
   potential. Experimental lattice parameters were employed for these
   calculations. The lattice thermal conductivities are plotted as a
   function of number of sampled phonon modes, i.e., product of number
   of sampled $q$-points and 3$n_\text{a}$, where $n_\text{a}=9$ for
   $\alpha$-quartz and $n_\text{a}=12$ for $\alpha$-cristobalite.
   \label{fig:q-convergence}}
  \end{center}
\end{figure}

In Fig.~\ref{fig:q-convergence}, convergences of lattice thermal
conductivities with respect to the number of sampling phonon modes in
Brillouin zones are presented. For both of $\alpha$-quartz and
$\alpha$-cristobalite, the lattice thermal
conductivities converge well by $\sim$10$^5$ phonon-mode sampling
points. Since we needed more sampling phonon modes to converge the curve
shapes of spectrum-like plots such as phonon density of states (DOS), we
chose the $19\times 19\times 19$ and $19\times 19\times 14$ $q$-point
sampling meshes for $\alpha$-quartz and $\alpha$-cristobalite,
respectively.

\begin{table}[ht]
 \caption{\label{table:kappa-lbte} Experimental and calculated lattice
 thermal conductivities $\kappa$ (W/m-K) of $\alpha$-quartz and
 $\alpha$-cristobalite at room temperature. In the calculations, RTA and
 direct solutions of linearized phonon Boltzmann transport equation are
 compared. Only the average values $\kappa_\text{av}$ were given in the
 report by Kunugi {\it et al.} For the calculations, we also show
 average values, here simply defined as
 $\kappa_\text{av}=(2\kappa_{xx}+\kappa_{zz})/3$, to make rough
 comparisons.}
 \begin{ruledtabular}
  \begin{tabular}{ccccc}
   & & $\kappa_{xx}$ & $\kappa_{zz}$ & $\kappa_\text{av}$ \\
   \hline
   \multirow{3}{*}{$\alpha$-quartz}
   & Exp. by Kanamori {\it et al.}\footnotemark[1] & 6.49 & 13.9 & - \\
   & Exp. by Kunugi {\it et al.}\footnotemark[2] & - & - & 7.15 \\
   & RTA & 4.9 & 10.7 & 6.8\\
   & Direct solution & 5.1 & 10.9 & 7.0 \\
   \hline
   \multirow{3}{*}{$\alpha$-cristobalite}
   & Exp. by Kunugi {\it et al.}\footnotemark[2] & - & - & 6.15 \\
   & RTA & 6.4 & 7.6 & 6.8 \\
   & Direct solution & 6.6 & 7.2 & 6.8 \\
  \end{tabular}
  \footnotetext[1]{Ref.~\onlinecite{Kanamori-1968}.}
  \footnotetext[2]{Ref.~\onlinecite{Kunugi-1972}.}
 \end{ruledtabular}
\end{table}

In Table~\ref{table:kappa-lbte}, experimental and calculated lattice
thermal conductivities are presented. For the calculations, we employed
RTA~\cite{Physics-of-phonons, phono3py} and
direct~\cite{Laurent-LBTE-2013, phono3py} solutions of linearized phonon
Boltzmann transport equation, and the obtained values were close each
other for $\alpha$-quartz and $\alpha$-cristobalite. Therefore, we
decided to use the RTA solution, since, compared with the direct
solution, it has an advantage in analyzing results more easily
and intuitively by its closed form of lattice thermal
conductivity formula.

Due to crystal symmetries of $\alpha$-quartz (trigonal) and
$\alpha$-cristobalite (tetragonal), both lattice thermal conductivity
tensors have only two degrees of freedom, $\kappa_{xx}$ and
$\kappa_{zz}$. $\alpha$-quartz exhibits largely anisotropic lattice
thermal conductivity whereas that of $\alpha$-cristobalite is more
isotropic as shown in Table~\ref{table:kappa-lbte}. From the
experimental measurement of $\alpha$-quartz by Kanamori {\it et
al.},\cite{Kanamori-1968} the ratio $\kappa_{zz}/\kappa_{xx}$ is around
2, which is well reproduced by our calculation. However each of
$\kappa_{xx}$ and $\kappa_{zz}$ from the calculation underestimates the
experimental values. There is another experimental measurement of powder
sample reported by Kunugi {\it et al.}\cite{Kunugi-1972} By taking
$\kappa_\text{av} = (2\kappa_{xx}+\kappa_{zz})/3$ as an averaged
value along orientations, the calculated value is found to be close to
the experiment. In the same report by Kunugi {\it et al.}, they also
showed the measurement of powder $\alpha$-cristobalite, which agrees well
with the averaged value by the present calculation.

\begin{table}[ht]
 \caption{\label{table:kappa-wrt-settings} Calculated lattice thermal
 conductivities $\kappa$ (W/m-K) of $\alpha$-quartz and
 $\alpha$-cristobalite at 300 K with respect to the choices of lattice
 parameters (see Table~\ref{table:latparams-wrt-settings}) and the
 exchange correlation potentials (XC-func.) of PBEsol and LDA.}
 \begin{ruledtabular}
  \begin{tabular}{lclccc}
   && lattice params. & XC-func. & $\kappa_{xx}$  & $\kappa_{zz}$ \\
   \hline
   \multirow{4}{*}{$\alpha$-quartz}
    && Calc./PBEsol & PBEsol & 4.2 & 8.7 \\
    && Calc./LDA & LDA & 4.9 & 10.8 \\
    && Exp. & PBEsol & 4.9 & 10.7 \\
    && Exp. & LDA & 4.3 & 9.2 \\
   \hline
   \multirow{4}{*}{$\alpha$-cristobalite}
    && Calc./PBEsol & PBEsol & 5.2 & 5.9 \\
    && Calc./LDA & LDA & 5.7 & 6.6 \\
    && Exp. & PBEsol & 6.4 & 7.6 \\
    && Exp. & LDA & 5.3 & 6.1 \\
  \end{tabular}
 \end{ruledtabular}
\end{table}

It is not always the case that we can fortunately refer to experimental
lattice parameters on lattice thermal conductivity
calculations. Therefore it is of interest to see how much different
lattice thermal conductivities are calculated using the
lattice parameters determined by the first-principles calculations and
those calculated with the experimental lattice parameters. The results
are shown in Table~\ref{table:kappa-wrt-settings}. For the calculated
lattice parameters by PBEsol (LDA) overestimated (underestimated) the
experimental lattice parameters as shown in Table
~\ref{table:latparams-wrt-settings}, smaller (larger) lattice thermal
conductivities were obtained following the general trend of the volume
dependence. When using the same experimental lattice parameters, the
lattice thermal conductivities calculated with PBEsol were obtained
larger than those with LDA for both $\alpha$-quartz and
$\alpha$-cristobalite. From these calculations, we can see
distinguishable effects by the choices of the exchange correlation
potentials: one is in determining lattice parameters and the other is in
calculating forces on atoms. However since the values and the ratios
$\kappa_{zz}/\kappa_{xx}$ in Table~\ref{table:kappa-wrt-settings} are
close enough, any choice given here is found a reasonable choice unless
we expect too good quantitative agreements between calculations and
experiments.

\begin{table}[ht]
 \caption{\label{table:kappa-wrt-other-settings} Calculated lattice
 thermal conductivities $\kappa$ (W/m-K) of $\alpha$-quartz and
 $\alpha$-cristobalite at 300 K with respect to supercell size used to
 calculate third-order force constants and plane-wave energy cutoff (eV)
 and atomic displacement distance used to calculate second- and
 third-order force constants.}
 \begin{ruledtabular}
  \begin{tabular}{cccccc}
   & supercell & displacement & PW cutoff & $\kappa_{xx}$  & $\kappa_{zz}$ \\
   \hline
   \multirow{8}{*}{$\alpha$-quartz} & $2\times 2\times 2$ & 0.03 & 520 & 4.9 & 10.7 \\
    & $3\times 3\times 2$ & 0.03 & 520 & 4.7 & 10.3 \\
    & $2\times 2\times 3$ & 0.03 & 520 & 4.7 & 10.5 \\
    & $1\times 1\times 1$ & 0.03 & 520 & 4.4 & 9.5 \\
    & $2\times 2\times 2$ & 0.03 & 600 & 5.0 & 10.8 \\
    & $2\times 2\times 2$ & 0.03 & 440 & 4.9 & 10.6 \\
    & $2\times 2\times 2$ & 0.01 & 520 & 5.1 & 11.0 \\
    & $2\times 2\times 2$ & 0.05 & 520 & 4.2 & 9.5 \\
   \hline
   \multirow{6}{*}{$\alpha$-cristobalite} & $2\times 2\times 2$ & 0.03 & 520 & 6.4 & 7.6 \\
    & $3\times 3\times 2$ & 0.03 & 520 & 6.3 & 7.4 \\
    & $2\times 2\times 1$ & 0.03 & 520 & 5.7 & 6.6 \\
    & $1\times 1\times 1$ & 0.03 & 520 & 5.0 & 5.3 \\
    & $2\times 2\times 2$ & 0.01 & 520 & 6.4 & 7.6 \\
    & $2\times 2\times 2$ & 0.05 & 520 & 6.2 & 7.4 \\
  \end{tabular}
 \end{ruledtabular}
\end{table}

We investigated the effects on calculated lattice thermal conductivities
by different choices of supercell size used for the calculation of the
third-order force constants and finite atomic displacement distance and
plane-wave cutoff energy used for the calculations of the second- and
third-order force constants. The $k$-points of the supercell reciprocal
spaces were sampled with equivalent density meshes to those written in
Sec.~\ref{sec:computational-details} except for that of $3\times 3\times
2$ supercell of $\alpha$-cristobalite where the $2\times 2\times 2$
sampling mesh shifted in half grid distances along all directions from
the $\Gamma$-point centered mesh was used. These results show, for both
$\alpha$-quartz and $\alpha$-cristobalite, that $2\times 2\times 2$
supercells are the reasonable choices considering the tradeoff of
convergences of the lattice thermal conductivity values and the required
computational demands (see Appendix) with respect to our current
computational resource. It also shows the use of the unit cells for
third-order force constants calculations is not a bad choice if a
purpose is the rough estimation.

The choice of 0.05 \AA~displacement distance induces decrease of lattice
thermal conductivity for $\alpha$-quartz. This is considered due to
inclusion of higher order anharmonicity. In general, decreasing the
displacement distance, numerical error in force constants is
magnified. The results by the choice of 0.01 \AA~displacement distance
give similar results with those by 0.03 \AA. This means that the
numerical errors and inclusions of higher order anharmonicity are
managed to be small by the choice of 0.03 \AA~displacement distance
for our computer simulation settings.

\section{Results and discussion}
 \label{sec:results}

 In RTA, lattice thermal conductivity $\kappa$ is written in a closed
form:\cite{Physics-of-phonons}
\begin{equation}
 \label{eq:kappa-rta}
 \kappa=\frac{1}{NV_0}\sum_\lambda C_\lambda \mathbf{v}_\lambda \otimes
  \mathbf{v}_\lambda \tau_\lambda,
\end{equation}
where $N$ and $V_0$ are the number of unit cells in the system and
volume of the unit cell, respectively. The suffix $\lambda$ represents
the phonon mode as the pair of phonon wave vector $\mathbf{q}$ and
branch $j$, $\lambda \equiv (\mathbf{q},j)$, and similarly we denote
$-\lambda \equiv (-\mathbf{q},j)$. $C_\lambda$ is the mode heat capacity
given as
\begin{equation}
C_\lambda = k_\mathrm{B}
 \left(\frac{\hbar\omega_\lambda}{k_\mathrm{B} T}\right)^2
 \frac{\exp(\hbar\omega_\lambda/k_\mathrm{B}
 T)}{[\exp(\hbar\omega_\lambda/k_\mathrm{B} T)-1]^2},
\end{equation}
where $\omega_\lambda = \omega(\mathbf{q}, j)$ is the phonon
frequency. $\mathbf{v}_\lambda$ is the phonon group velocity defined as
the gradient of the phonon energy surface:
\begin{equation}
 \mathbf{v}_\lambda = \nabla_\mathbf{q} \omega(\mathbf{q}, j).
\end{equation}
$\tau_\lambda$ is the single-mode relaxation time and we use phonon
lifetime as $\tau_\lambda$. We calculated phonon lifetime
$\tau_\lambda=\frac{1}{2\Gamma_\lambda(\omega_\lambda)}$
by~\cite{Laurent-phph-2011, phono3py}
\begin{align}
 \label{eq:gamma}
 \Gamma_\lambda(\omega) = \frac{18\pi}{\hbar^2}
  \sum_{\lambda' \lambda''} \Delta(- \mathbf{q} + \mathbf{q}' + \mathbf{q}'')
  N_{\lambda'\lambda''}(\omega) \bigl|\Phi_{-\lambda\lambda'\lambda''}\bigl|^2,
\end{align}
where
\begin{align}
N_{\lambda'\lambda''}(\omega) &= (n_{\lambda'}+ n_{\lambda''}+1)
 \delta( \omega - \omega_{\lambda'} - \omega_{\lambda''}) + \nonumber \\
 (n_{\lambda'} - & n_{\lambda''})
 [ \delta( \omega + \omega_{\lambda'} -
 \omega_{\lambda''}) - \delta(  \omega - \omega_{\lambda'} +
 \omega_{\lambda''})],
\end{align}
with $n_\lambda = [\exp(\hbar\omega_\lambda/k_\mathrm{B}T) - 1]^{-1}$ as
the phonon occupation number at equilibrium. In these equations, $\hbar$
and $k_\mathrm{B}$ denote the reduced Planck constant and Boltzmann
constant, respectively. $\Phi_{\lambda\lambda'\lambda''}$ gives the
phonon-phonon interaction strength among three phonons calculated from
second- and third-order force constants. $\Delta(\mathbf{q} +
\mathbf{q}' + \mathbf{q}'') \equiv 1$ if $\mathbf{q} + \mathbf{q}' +
\mathbf{q}''=\mathbf{G}$ otherwise 0, where $\mathbf{G}$ is the
reciprocal lattice vector. This constraint comes from the lattice
translational invariance that appears inside
$\Phi_{\lambda\lambda'\lambda''}$,\cite{phono3py} however it is made
visible in Eq.~(\ref{eq:gamma}) for the analysis given below. More
details such as the phase convention, coefficients, and
$\Phi_{\lambda\lambda'\lambda''}$ are found in
Ref.~\onlinecite{phono3py}.

\begin{figure*}[ht]
  \begin{center}
   \includegraphics[width=0.90\linewidth]{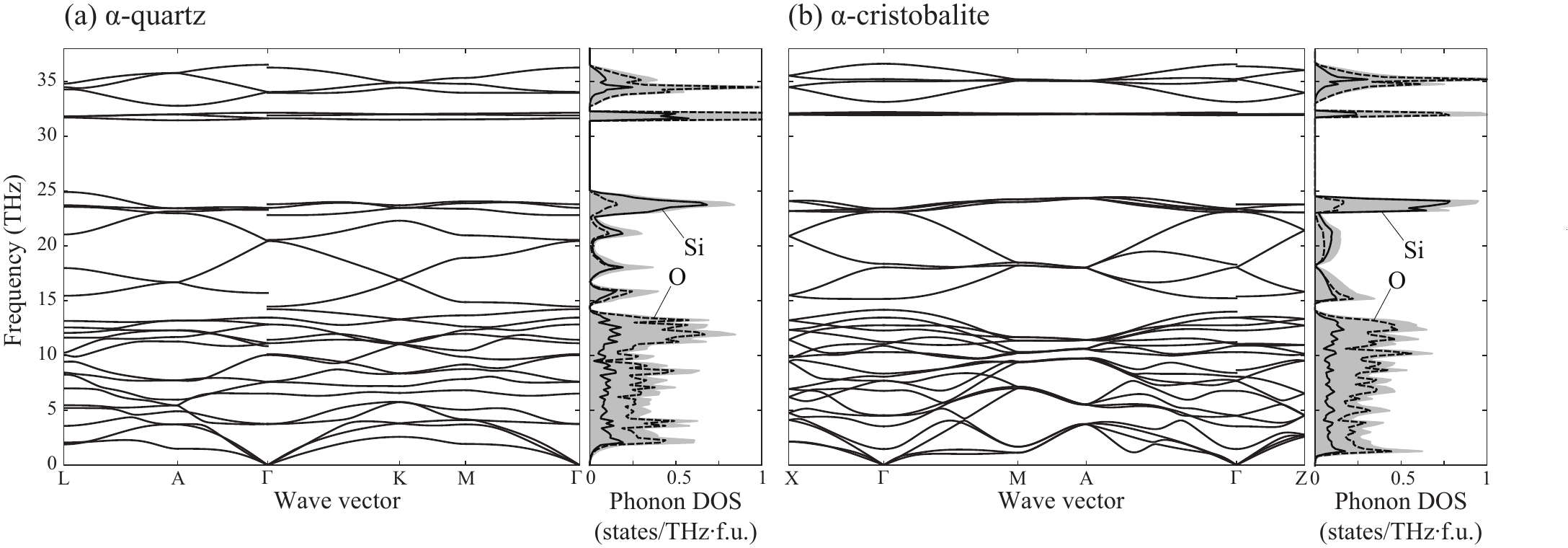}
   \caption{Phonon band structures and DOS of (a) $\alpha$-quartz and
   (b) $\alpha$-cristobalite. In the DOS plots on the right hand sides
   of the band structures, the solid and dotted curves depict the partial
   DOS of Si and O, respectively, and the curves under shadow show the
   total DOS. The special point symbols of wave vectors follow the
   convention provided in the Bilbao crystallographic
   server.\cite{Aroyo-2014}} \label{fig:bandstructures}
  \end{center}
\end{figure*}

Phonon band structures and DOS of $\alpha$-quartz and
$\alpha$-cristobalite are shown in Figs.~\ref{fig:bandstructures} (a)
and (b), respectively. These phonon structures in their shapes show
reasonable agreements with previous calculations and experiments
reported in Refs.~\onlinecite{Dorner-1980, Strauch-1993, Gonze-1994,
Swainson-1995, Dove-1997, Wehinger-2015}. Between $\alpha$-quartz and
$\alpha$-cristobalite, their total and partial DOS curves are analogous.
In detail, the position of the first peak of $\alpha$-quartz from 0 THz
is located at higher phonon frequency than that of
$\alpha$-cristobalite. Their first peak positions roughly correspond to
M- and L-points of $\alpha$-quartz and M-point of $\alpha$-cristobalite
in respective phonon band structures. These low phonon modes are
considered to be made of rigid unit motions of SiO$_4$
tetrahedra,\cite{Giddy-1993,Dove-2000,Wells-2002} i.e., the phonon band
structures at low frequencies reflect the different styles of the
tetrahedron linkages.

\begin{figure}[ht]
  \begin{center}
   \includegraphics[width=1.00\linewidth]{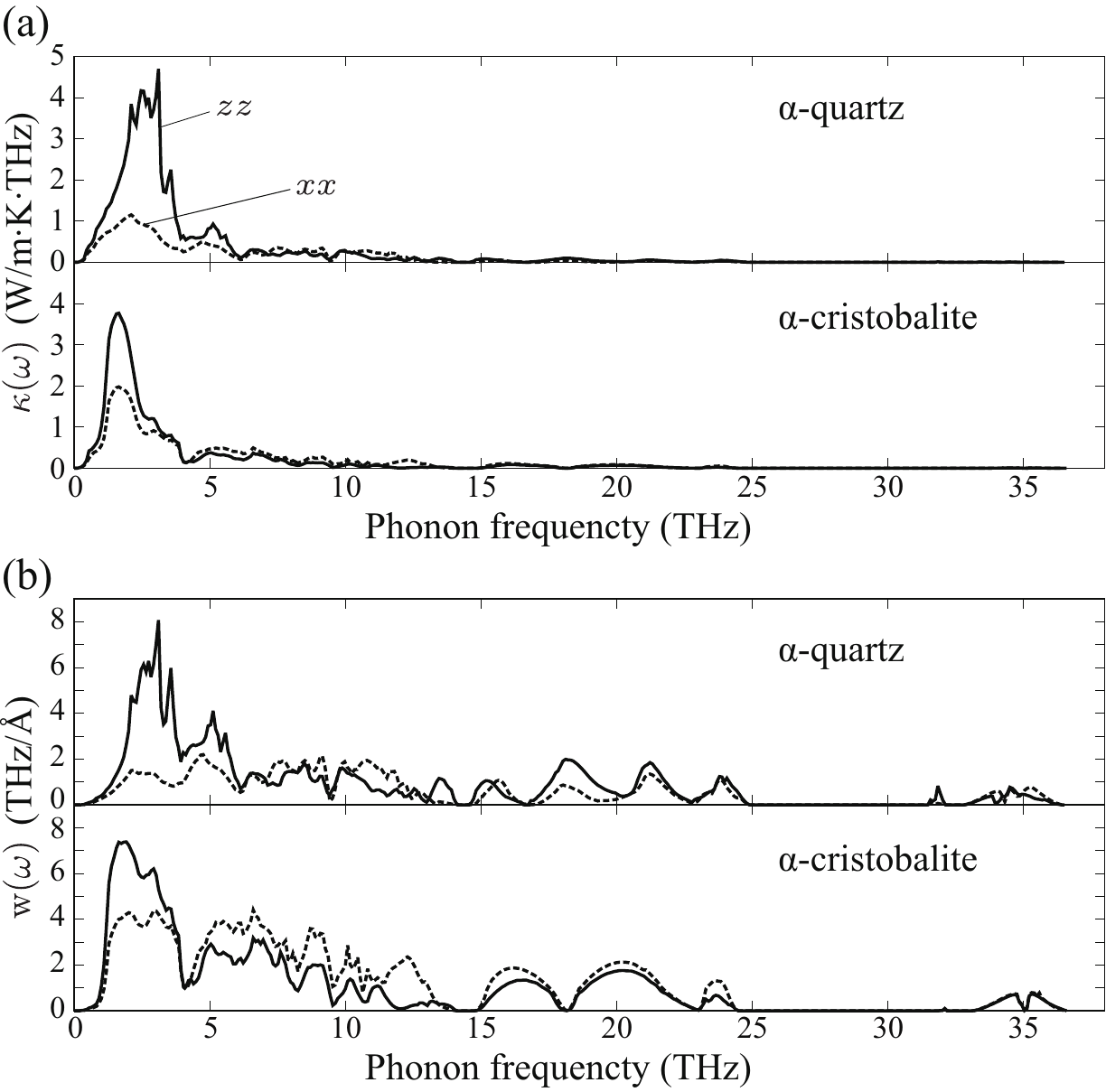} \caption{ (a)
   Densities of lattice thermal conductivities $\kappa(\omega)$ at 300K
   and (b) distributions of direct-vector-products of group velocities
   $\mathrm{w}(\omega)$ (see Eq.~(\ref{eq:gv-density})) calculated for
   $\alpha$-quartz and $\alpha$-cristobalite with respect to phonon
   frequency. Both in (a) and (b), dotted and solid curves depict their
   $xx$ and $zz$ components, respectively. \label{fig:dkaccum}}
  \end{center}
\end{figure}

To visualize phonon mode contribution to lattice thermal conductivity,
we define densities of lattice thermal conductivities $\kappa(\omega)$ as
\begin{equation}
 \label{eq:kappa-density}
  \kappa = \int_0^\infty \kappa(\omega) d\omega
\end{equation}
with
\begin{align}
 \label{eq:kappa-density-rta}
 \kappa(\omega) &\equiv \frac{1}{NV_0}\sum_{\lambda} C_{\lambda}
 \mathbf{v}_{\lambda} \otimes
  \mathbf{v}_{\lambda} \tau_{\lambda} \delta(\omega - \omega_\lambda).
\end{align}
Compared with phonon DOS written as $1/N \sum_\lambda \delta(\omega
-\omega_\lambda)$, Eq.~(\ref{eq:kappa-density-rta}) is considered as a
weighted DOS and each weight $C_{\lambda} \mathbf{v}_{\lambda} \otimes
\mathbf{v}_{\lambda} \tau_{\lambda} /V_0$ is understood as microscopic
contribution of phonon mode $\lambda$ to lattice thermal conductivity at
$\omega_\lambda$. In Fig.~\ref{fig:dkaccum} (a), $\kappa(\omega)$ of
$\alpha$-quartz and $\alpha$-cristobalite are drawn as a function of
phonon frequency. We can see large peaks below 5 THz, where the phonon
modes determine more than halves of $\kappa_{xx}$ and $\kappa_{zz}$ of
$\alpha$-quartz and $\alpha$-cristobalite. The curve shapes of
$\kappa(\omega)$ are similar to those of the phonon DOS below their
first peaks. Therefore it is considered that the number of states is the
dominating factor of the lattice thermal conductivities in these phonon
frequency ranges. Above 5 THz, $\kappa(\omega)$ are relatively small,
however they contribute little by little to $\kappa$ up to $\sim$25 THz.

\begin{figure}[ht]
  \begin{center}
   \includegraphics[width=1.00\linewidth]{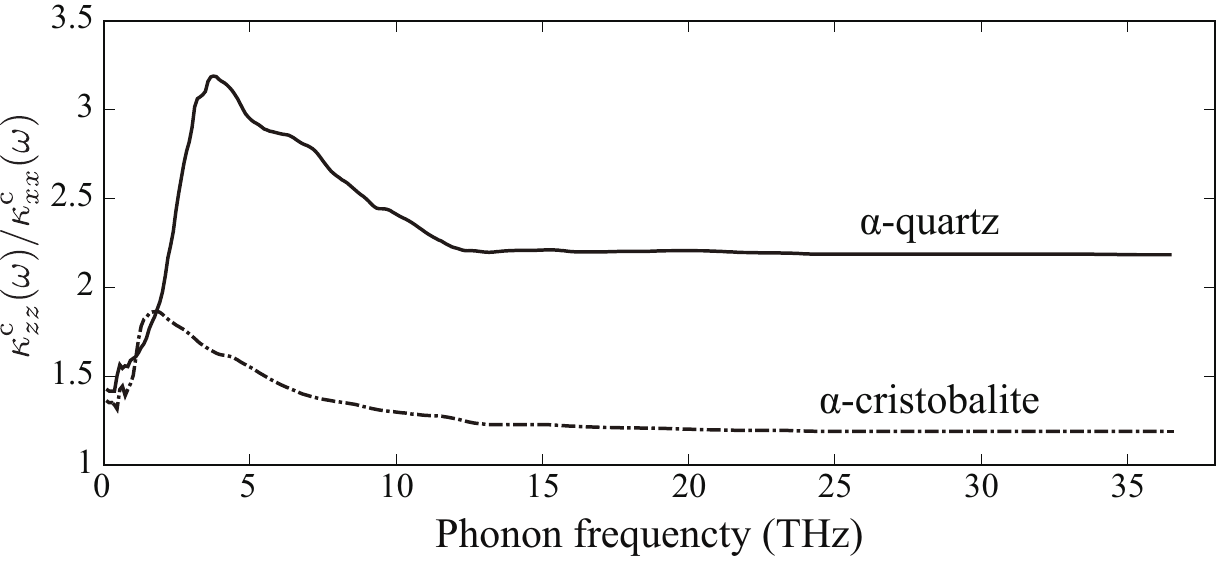}
   \caption{Ratios between $zz$ and $xx$ elements of cumulative lattice
   thermal conductivities,
   $\kappa^\text{c}_{zz}(\omega)/\kappa^\text{c}_{xx}(\omega)$, in
   $\alpha$-quartz (solid curve) and $\alpha$-cristobalite
   (dashed-dotted curve) at 300 K. \label{fig:kappa-ratio}}
  \end{center}
\end{figure}

Anisotropy of lattice thermal conductivity, i.e., the ratio
$\kappa_{zz}/\kappa_{xx}$, is larger in $\alpha$-quartz than in
$\alpha$-cristobalite. The phonon mode contributions to the anisotropic
$\kappa$ are discussed using cumulative lattice thermal conductivity
given by
\begin{equation}
  \kappa^\text{c}(\omega) = \int_0^\omega \kappa(\omega') d\omega'.
\end{equation}
Obviously $\lim_{\omega\rightarrow\infty}\kappa^\text{c}(\omega) =
\kappa$ from Eq.~(\ref{eq:kappa-density}). The ratios
$\kappa^\text{c}_{zz}(\omega)/\kappa^\text{c}_{xx}(\omega)$ are shown in
Fig.~\ref{fig:kappa-ratio}, where $\alpha$-quartz and
$\alpha$-cristobalite present similar behaviours, although their
intensities are different. Increasing phonon frequency from 0 THz, their
ratios rapidly increase at low phonon frequencies and start to decrease
gently until the ratios become $\kappa_{zz}/\kappa_{xx}$.  This
difference is exhibited in distributions of $\mathbf{v}_{\lambda}
\otimes \mathbf{v}_{\lambda}$ that are written in analogy to
$\kappa(\omega)$ of Eq.~(\ref{eq:kappa-density-rta}) as
\begin{align}
 \label{eq:gv-density}
 \mathrm{w}(\omega) &\equiv \frac{1}{NV_0}\sum_{\lambda}
 \mathbf{v}_{\lambda} \otimes
  \mathbf{v}_{\lambda} \delta(\omega - \omega_\lambda).
\end{align}
$\mathrm{w}(\omega)$ are shown in Fig.~\ref{fig:dkaccum} (b). Below 5
THz, the ratio between $\mathrm{w}_{zz}(\omega)$ and
$\mathrm{w}_{xx}(\omega)$ is clearly larger in $\alpha$-quartz
than in $\alpha$-cristobalite.

\begin{figure}[ht]
  \begin{center}
   \includegraphics[width=1.00\linewidth]{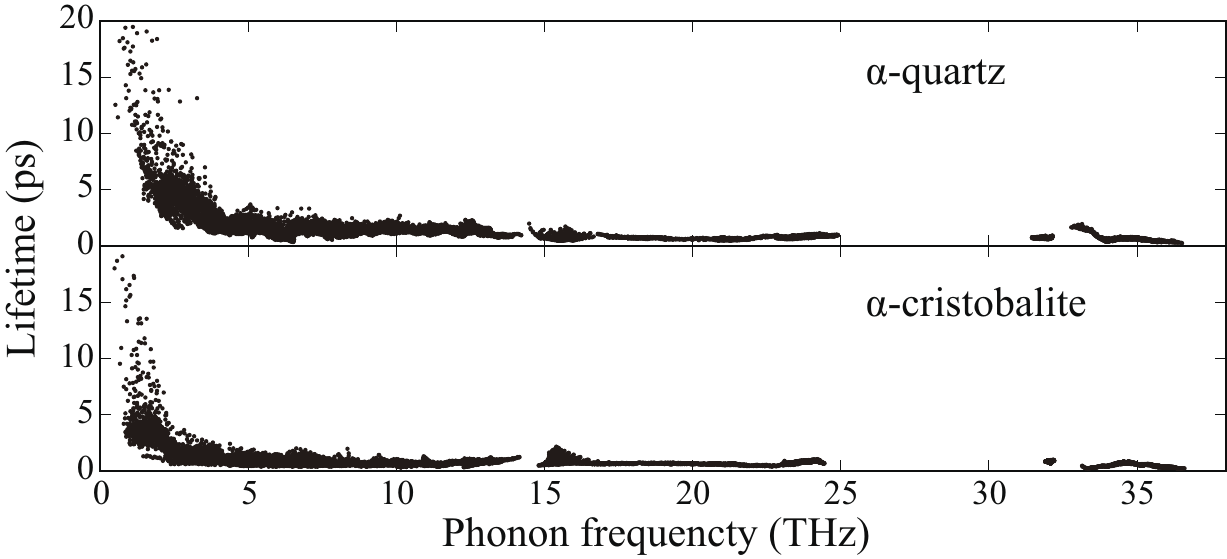} \caption{ Phonon
   lifetimes of $\alpha$-quartz and $\alpha$-cristobalite at 300K with
   respect to phonon frequency. Each dot corresponds to one phonon
   mode. The points are sampled on the $19\times 19\times 19$ mesh for
   $\alpha$-quartz and $19\times 19\times 14$ mesh for
   $\alpha$-cristobalite in the respective Brillouin
   zones. \label{fig:tau}}
  \end{center}
\end{figure}

Comparing Figs.~\ref{fig:dkaccum} (a) and (b), increasing phonon
frequency, $\kappa(\omega)$ more quickly decrease after first large
peaks than $\mathrm{w}(\omega)$ in both $\alpha$-quartz and
$\alpha$-cristobalite. This is due to phonon frequency dependencies of
$C_\lambda\tau_{\lambda}$, however the effect of $C_\lambda$ to the
curve shapes of $\kappa(\omega)$ with respect to those of
$\mathrm{w}(\omega)$ is small since $C_\lambda$ is approximately
constant ${\sim}k_\mathrm{B}$ at 300 K below 10 Hz. In
Fig.~\ref{fig:tau}, $\tau_\lambda$ are plotted by dots as a function of
phonon frequency. At lower phonon frequencies, phonons tend to have
longer lifetimes and decrease quickly their lifetimes increasing phonon
frequency below 5 THz. Both of $\alpha$-quartz and $\alpha$-cristobalite
show the same trend but with different rate of decrease, which clearly
impacts to the shapes of $\kappa(\omega)$ in Fig.~\ref{fig:dkaccum} (a),
e.g., $\kappa(\omega)$ of $\alpha$-cristobalite corresponding to the
second peak of $\mathrm{w}(\omega)$ at $\sim$3 THz is removed by the
decrease of $\tau_\lambda$.

\begin{figure}[ht]
  \begin{center}
   \includegraphics[width=1.00\linewidth]{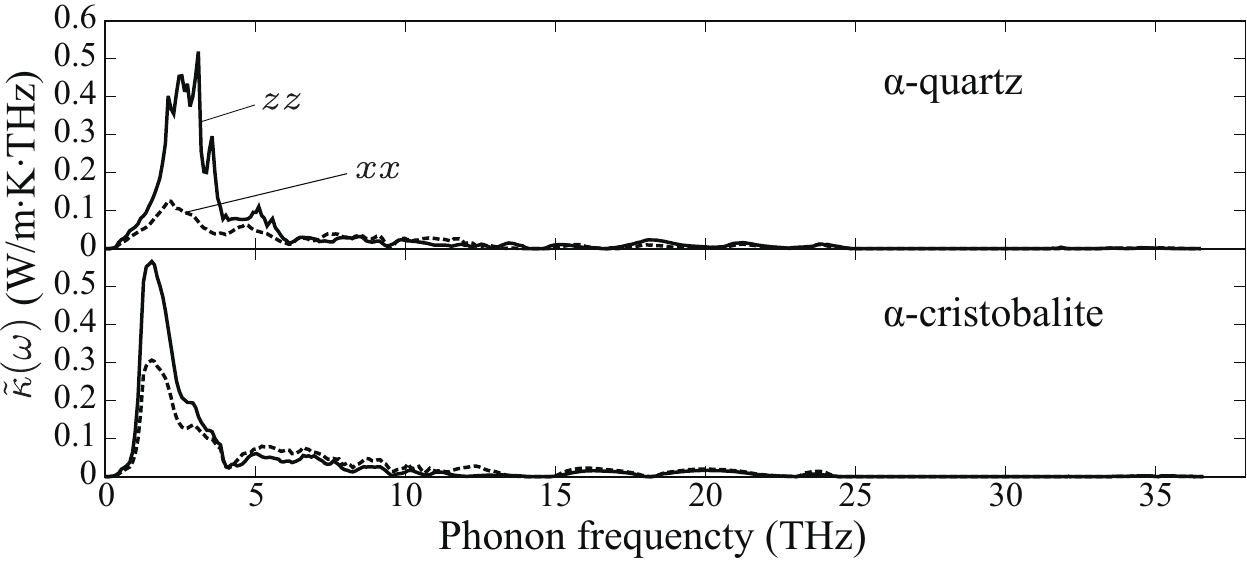}
   \caption{$\tilde{\kappa}(\omega)$, densities of lattice thermal
   conductivities of $\alpha$-quartz and $\alpha$-cristobalite
   calculated with $\tilde{P}=\tilde{P}_\text{av}$ at 300 K as a
   function of phonon frequency. Dotted and solid curves depict
   $\tilde{\kappa}_{xx}$ and $\tilde{\kappa}_{zz}$,
   respectively. \label{fig:dkaccum-ave}}
  \end{center}
\end{figure}

\begin{table}[ht]
 \caption{\label{table:kappa-tilde} $\tilde{P}_\text{av}$ (meV$^2$) (see
 Eq.~(\ref{eq:P-ave})) and lattice thermal conductivities
 $\tilde{\kappa}$ (W/m-K) of $\alpha$-quartz and $\alpha$-cristobalite
 calculated using $\tilde{P}_\text{av}$ at 300K. To align the scale of
 $\tilde{P}_\text{av}$ between $\alpha$-quartz and $\alpha$-cristobalite,
 $(3n_\text{a})^2N$ are multiplied with corresponding $\tilde{P}_\text{av}$,
 where $n_\text{a}=9$ and $N=19\times19\times19$ for
 $\alpha$-quartz and $n_\text{a}=12$ and $N=19\times19\times14$ for
 $\alpha$-cristobalite.}
 \begin{ruledtabular}
  \begin{tabular}{cccc}
   & $(3n_\text{a})^2N\tilde{P}_\text{av}$  & $\tilde{\kappa}_{xx}$
   & $\tilde{\kappa}_{zz}$ \\
   \hline
   $\alpha$-quartz & 2.67 & 0.54 & 1.16 \\
   $\alpha$-cristobalite & 2.78 & 1.00 & 1.17 \\
  \end{tabular}
 \end{ruledtabular}
\end{table}

Recalling Eq.~(\ref{eq:gamma}), $\tau_\lambda$ is constructed from the
wave vector constraint $\Delta(\mathbf{q}+\mathbf{q}'+\mathbf{q}'')$,
weighted energy conservation $N_{\lambda'\lambda''}(\omega_\lambda)$,
and $|\Phi_{\lambda \lambda' \lambda''}|^2$. To make our discussion
simple, we replace $|\Phi_{\lambda \lambda' \lambda''}|^2$ by a constant
value $\tilde{P}$ if $\mathbf{q}+\mathbf{q}'+\mathbf{q}''=\mathbf{G}$ or
by 0 otherwise. As an attempt, we use $\tilde{P}=\tilde{P}_\text{av}$
defined as an average of $|\Phi_{\lambda \lambda' \lambda''}|^2$ by
\begin{equation}
 \label{eq:P-ave}
\tilde{P}_\text{av} \equiv
\frac{1}{(3n_\text{a})^3N^2}\sum_{\lambda\lambda' \lambda''}
\bigl|\Phi_{\lambda \lambda' \lambda''}\bigl|^2
= \frac{1}{3n_\text{a}N^2}\sum_\lambda P_\lambda,
\end{equation}
where $P_\lambda$ is that for one phonon mode:\cite{phono3py}
\begin{equation}
 \label{eq:P-ave-one}
 P_\lambda \equiv \frac{1}{(3n_\text{a})^2}\sum_{\lambda'
\lambda''}|\Phi_{\lambda \lambda' \lambda''}|^2.
\end{equation}
Since $(3n_\text{a})^2N\tilde{P}_\text{av}$ of $\alpha$-quartz
and $\alpha$-cristobalite give the equivalent values as shown in
Table.\ref{table:kappa-tilde}, we expect that they have
similar phonon-phonon interaction strengths.

With $\tilde{P}$, ${\Gamma}_\lambda(\omega)$ is reduced to
\begin{align}
 \label{eq:gamma-ave}
 \tilde{\Gamma}_\lambda(\omega) = \frac{18\pi}{\hbar^2}  \tilde{P}
  \sum_{\lambda' \lambda''} \Delta(- \mathbf{q} + \mathbf{q}' + \mathbf{q}'')
  N_{\lambda'\lambda''}(\omega).
\end{align}
In Eq.~(\ref{eq:gamma-ave}), the summation on the right hand side is
made of three phonon scattering channels weighted by phonon occupation
numbers, which can be computed from the second-order force constants.
The lattice thermal conductivities calculated with
$\tilde{P}=\tilde{P}_\text{av}$ and
$\tilde{\tau}_\lambda=(2\tilde{\Gamma}_\lambda)^{-1}$, that we denote
$\tilde{\kappa}$, are presented in Table~\ref{table:kappa-tilde}. These
values are one order of magnitude smaller than the values in
Table~\ref{table:kappa-lbte}, however the anisotropies
$\tilde{\kappa}_{zz}/\tilde{\kappa}_{xx}$ are well reproduced, and as
shown in Fig.~\ref{fig:dkaccum-ave}, the curve shapes of the densities
of lattice thermal conductivities, denoted by $\tilde{\kappa}(\omega)$,
are almost identical to those of $\kappa(\omega)$ presented in
Fig.~\ref{fig:dkaccum} (a).

\begin{figure}[ht]
  \begin{center}
   \includegraphics[width=1.00\linewidth]{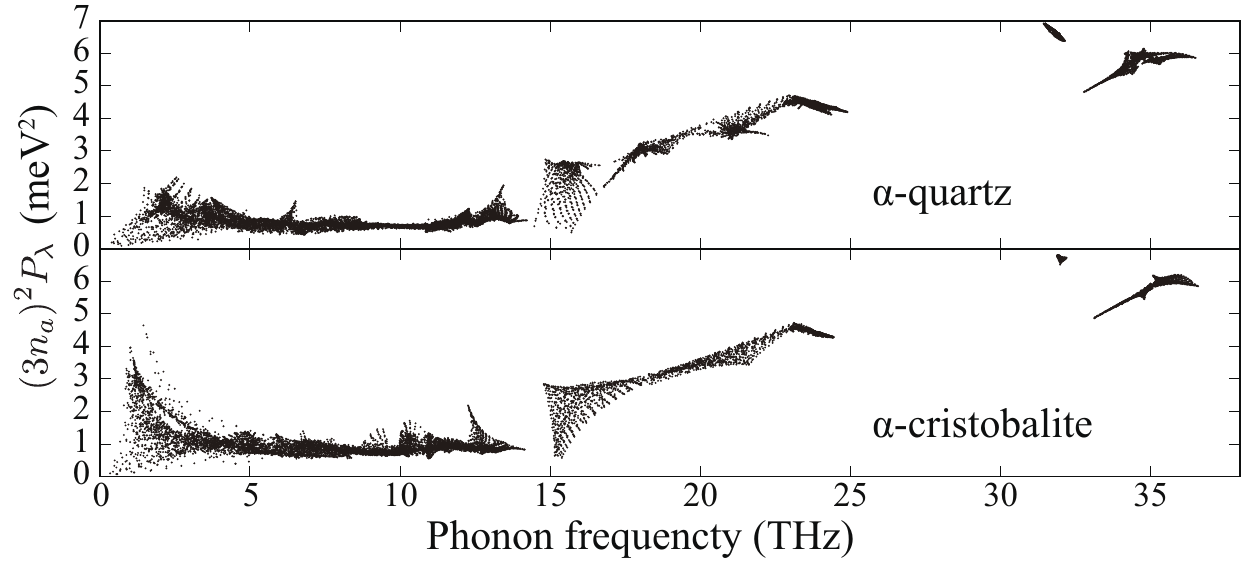} \caption{
   $(3n_\mathrm{a})^2P_\lambda$ of $\alpha$-quartz and
   $\alpha$-cristobalite with respect to phonon frequency. Here
   $(3n_\mathrm{a})^2$ is multiplied with $P_\lambda$ to align the scale
   between $\alpha$-quartz and $\alpha$-cristobalite. Each dot
   corresponds to one phonon mode. The points are sampled on the
   $19\times 19\times 19$ mesh for $\alpha$-quartz and $19\times
   19\times 14$ mesh for $\alpha$-cristobalite in the respective
   Brillouin zones.  \label{fig:ave-pp}}
  \end{center}
\end{figure}

\begin{figure}[ht]
  \begin{center}
   \includegraphics[width=1.00\linewidth]{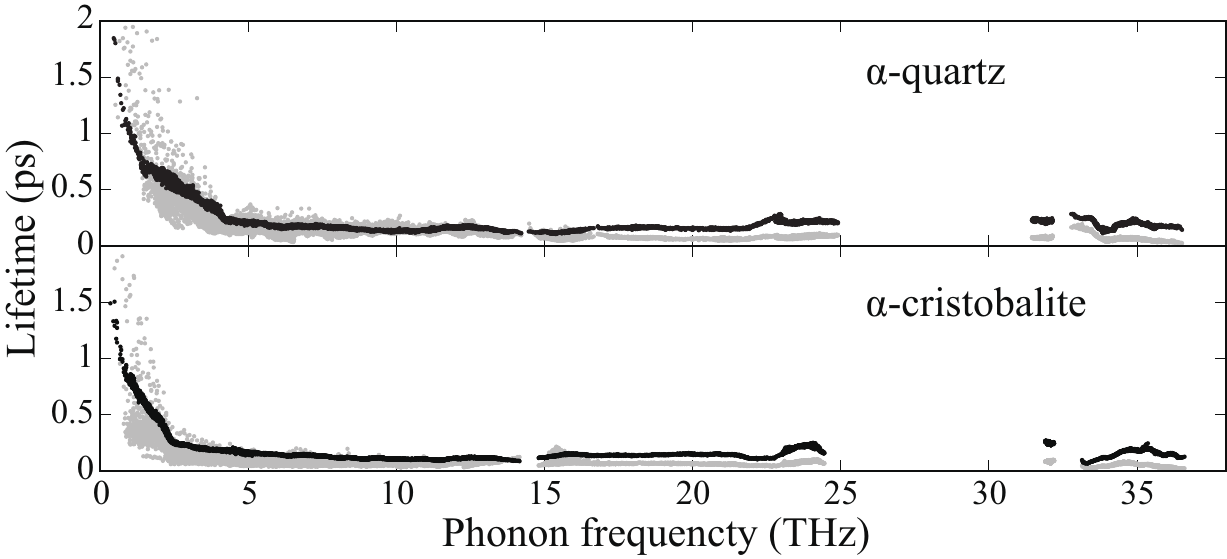} \caption{
   $\tilde{\tau}_\lambda$, phonon lifetimes of $\alpha$-quartz and
   $\alpha$-cristobalite calculated with $\tilde{P}=\tilde{P}_\text{av}$
   at 300 K as a function of phonon frequency (black dots). To compare,
   $\tau_\lambda \times 10^{-1}$ (Fig.~\ref{fig:tau}) are shown as the
   grey dots behind the black dots. The points are sampled on the
   $19\times 19\times 19$ mesh for $\alpha$-quartz and $19\times
   19\times 14$ mesh for $\alpha$-cristobalite in the respective
   Brillouin zones.  \label{fig:tau-ave}}
  \end{center}
\end{figure}

In Fig.~\ref{fig:ave-pp}, $P_\lambda$ of $\alpha$-quartz and
$\alpha$-cristobalite are plotted as a function of phonon
frequency. Their distributions are similar except at low phonon
frequency domains where the phonon DOS are small, which indicates that
the different styles of linkages of SiO$_4$ tetrahedra in their crystal
structures impact little to determine the phonon-phonon interaction
strengths. In Fig.~\ref{fig:ave-pp}, there are four characteristic
phonon frequency domains. The locations of these domains synchronize
with the phonon DOS shown in Figs.~\ref{fig:bandstructures} (a) and
(b). Between 0 to 15 THz, $P_\lambda$ are roughly constant, by which,
apart from their different magnitudes, similar phonon frequency
dependencies of $\tilde{\tau}_\lambda$ to those of $\tau_\lambda$ are
obtained as shown in Fig.~\ref{fig:tau-ave}. This enables the curve
shapes of $\tilde{\kappa}(\omega)$ to become equivalent to those of
$\kappa(\omega)$. Since more than 90\% of the lattice thermal
conductivities of $\alpha$-quartz and $\alpha$-cristobalite are
recovered in $\kappa(\omega)$ below 15 THz, having a good estimate of
the constant value, e.g., $\tilde{P} \sim \tilde{P}_\text{av}\times
10^{-1}$, it is considered possible to predict the lattice thermal
conductivities without computing third order force
constants. $P_\lambda$ start to increase from $\sim$15 THz to the phonon
band gap at $\sim$25 THz. The two small domains above 30 THz correspond
to respective two localized phonon bands. The ratio of Si and O partial
DOS gradually increases by increasing phonon frequency below 15
THz. This represents that SiO$_4$ rigid units vibrate translationally at
lower phonon frequencies and rotationally at increasing phonon
frequencies. Above 15 THz, it is considered that the larger $P_\lambda$,
i.e., larger anharmonicity, arises due to phonons that distort SiO$_4$
tetrahedron units.

\section{Summary}
The lattice thermal conductivity calculations were performed for
$\alpha$-quartz and $\alpha$-cristobalite using first-principles
anharmonic phonon calculation and linearized phonon Boltzmann transport
equation. Since direct and RTA solutions gave similar values of the
lattice thermal conductivities that also agree well with the
experimental values, we focused on our discussion using the RTA
solutions and phonon frequency dependencies of the phonon
properties. The densities of the lattice thermal conductivities
$\kappa(\omega)$ show the characteristic differences of phonon mode
contributions to the lattice thermal conductivities between
$\alpha$-quartz and $\alpha$-cristobalite. Below 2 THz for
$\alpha$-cristobalite and 3 THz for $\alpha$-quartz, phonon DOS and
$\mathbf{v}_\lambda \otimes \mathbf{v}_\lambda$ determines the shapes of
$\kappa(\omega)$. Above 5 THz, $\kappa(\omega)$ becomes much smaller
than those below 5 THz following the phonon frequency dependence of
$\tau_\lambda$. The large difference of anisotropies in the lattice
thermal conductivities of $\alpha$-quartz and $\alpha$-cristobalite was
found. This is mainly attributed by the distributions of the phonon
group velocities below 5 THz. The distributions of the phonon lifetimes
effective to determine the lattice thermal conductivities around room
temperature were well described by the momentum conservation
$\Delta(\mathbf{q} + \mathbf{q}' + \mathbf{q}'')$, the energy
conservation weighted by the phonon occupation numbers
$N_{\lambda'\lambda''}(\omega_\lambda)$, and a parameter $\tilde{P}$
that represents the phonon-phonon interaction strengths.

\section*{ACKNOWLEDGMENTS}
This work was supported by Grant-in-Aid for Scientific Research on
Innovative Areas ``Nano Informatics'' (Grant No. 25106005) from the
Japan Society for the Promotion of Science (JSPS), by MEXT Japan through
ESISM (Elements Strategy Initiative for Structural Materials) of Kyoto
University, and by the ``Materials Research by Information Integration''
Initiative (MI$^2$I) of the JST.

\appendix

\section{Effect of using real-space cutoff to calculate supercell
 third-order force constants}

Use of real-space cutoff distance to compute third-order force constants
in the supercell approach may drastically reduce computational demand of
lattice thermal conductivity calculation. However it should be used
carefully since the side effect such as degradation of the numerical
quality has not been well understood. In this Appendix, we provide our
examinations on the effect of using a cutoff distance for the
third-order force constants calculations. There are many possible ways
to cut off third-order force constants. Below, we explain our scheme and
show the convergence analysis.

\subsection{Scheme}

We calculate supercell third-order force constant element from two
atomic displacements and a force on an atom by,\cite{phono3py}
\begin{equation}
 \Phi_{\alpha\beta\gamma}(l\kappa,l'\kappa',l''\kappa'')
 \simeq -\frac{F_\gamma[l''\kappa''; \mathbf{u}(l\kappa),
  \mathbf{u}(l'\kappa')]}{u_\alpha(l\kappa)
  u_\beta(l'\kappa')},
\end{equation}
where $u_\alpha(l\kappa)$ means the finite displacement of the atom at
the position $\mathbf{r}(l\kappa)$ along $\alpha$-th Cartesian axis. The
indices $l$ and $\kappa$ denote the lattice point and the atom in the
unit cell, respectively. $F_\gamma[l''\kappa'';
\mathbf{u}(l\kappa),\mathbf{u}(l'\kappa')]$ gives the force that the atom
$l''\kappa''$ experiences by two atomic displacements
$\mathbf{u}(l\kappa)$ and $\mathbf{u}(l'\kappa')$. Here it is assumed
that we can obtain forces on all atoms in the supercell at once by each
supercell calculation with a pair of atomic displacements. This
assumption is currently normal in the DFT calculations since the
computation of forces from existing electronic wave function requires
relatively small computation.

Our cutoff distance $R_\text{cut}$ is used to collect all the displaced
atomic pairs whose distances
$\sqrt{|\mathbf{r}(l\kappa)-\mathbf{r}(l'\kappa')|^2}$ are shorter than
$R_\text{cut}$. The set of these pair displacements fills all supercell
third-order force constant elements except for the elements whose three
atoms are mutually more distant than $R_\text{cut}$.

\subsection{Results}

\begin{figure*}[ht]
  \begin{center}
   \includegraphics[width=0.90\linewidth]{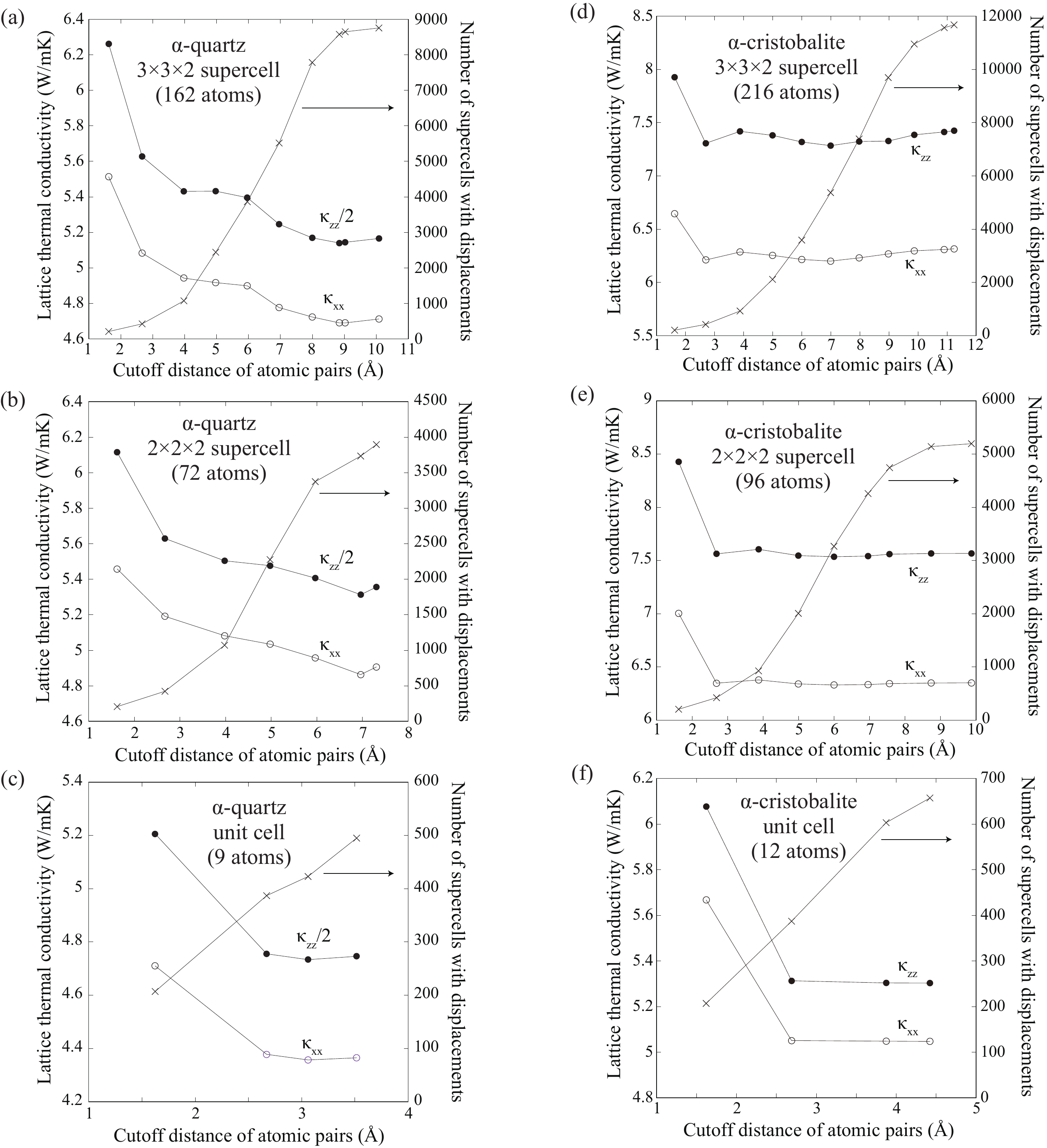}
   \caption{Lattice thermal conductivities at 300 K with respect to
   cutoff distances of atomic pairs used to compute third-order force
   constants employing (a) $\alpha$-quartz $3\times 3\times 2$ supercell
   (162 atoms), (b) $\alpha$-quartz $2\times 2\times 2$ supercell (72
   atoms), (c) $\alpha$-quartz unit cell (9 atoms), (d)
   $\alpha$-cristobalite $3\times 3\times 2$ supercell (216 atoms), (e)
   $\alpha$-cristobalite $2\times 2\times 2$ supercell (96 atoms), and
   (f) $\alpha$-cristobalite unit cell (12 atoms). The selected cutoff
   distances are those closest to but below (a) $2, \ldots, 11$ \AA, (b)
   $2,\ldots,8$ \AA, and (c) 2, 3, 3.5, and 4 \AA, (d) $2,\ldots,12$
   \AA, (e) $2,\ldots,10$ \AA, and (f) $2,\ldots,5$ \AA,
   respectively. The filled circles depict $\kappa_{zz}/2$ for
   $\alpha$-quartz and $\kappa_{zz}$ for $\alpha$-cristobalite, and the
   open circles show $\kappa_{xx}$. The cross symbols present the
   numbers of supercells with displacements that were used to compute
   the third-order force constants with the respective cutoff
   distances. The rightmost points correspond to the results obtained
   without using the cutoff distances. Lines are eye guides.}
   \label{fig:cutoff-pair-convergence}
  \end{center}
\end{figure*}

In this section, we present calculated lattice thermal conductivities
using different cutoff distances and see the convergences in
$\alpha$-quartz and $\alpha$-cristobalite using different supercell
sizes. We employed $3\times 3\times 2$ and $2\times 2\times 2$
supercells and unit cells for these examinations. The computations of
third-order force constants using the $3\times 3\times 2$ supercells
were computationally very demanding for our current computational
resource to fill all the elements, but not with the $2\times 2\times 2$.
For $\alpha$-quartz, one $3\times 3\times 2$ supercell calculation was
five times more computationally demanding than one $2\times 2\times 2$
supercell calculation. For $\alpha$-cristobalite, that was nine times
because of the denser $k$-point sampling for the $3\times
3\times 2$ supercell calculation of $\alpha$-cristobalite.

The purpose to use the cutoff distance is to obtain accurate third-order
force constants with reasonable computational demand though it is
safer to compute all elements of supercell force constants to
avoid sudden cut of those elements since it is difficult to predict what
happens after Fourier transformation of the third-order force constants
with the cut.

In Figs.~\ref{fig:cutoff-pair-convergence} (a), (b), and (c), the
lattice thermal conductivities of $\alpha$-quartz calculated against the
cutoff distances are shown for three different supercell sizes. The
lattice thermal conductivities generally decrease increasing the cutoff
distance in these supercell sizes. It looks that each lattice thermal
conductivity converges toward its rightmost point that corresponds to
the full calculation where all elements of the supercell third-order
force constants were filled.  In Fig.~\ref{fig:cutoff-pair-convergence}
(b), at the rightmost point, the lattice thermal conductivity increases
in contradiction to the tendency of decreasing with increasing the
cutoff distance. This is considered a visible effect of the cut
of the supercell third-order force constants elements.
For $\alpha$-cristobalite as shown in
Figs.~\ref{fig:cutoff-pair-convergence} (d), (e), and (f), the
convergence is achieved at relatively shorter cutoff distance of $\sim$4
\AA. This is about the distance between two atoms in neighboring SiO$_4$
tetrahedra. However the calculation of the third-order force constants
with the $3\times 3\times 2$ supercell using $\sim$4 \AA~cutoff distance
is already more computationally demanding than the full calculation with
the $2\times 2\times 2$ supercell. Therefore the supercell size has to
be chosen systematically along with the choice of the cutoff distance.
Comparing Figs.~\ref{fig:cutoff-pair-convergence} (a) and (d), we can
see lattice thermal conductivity of $\alpha$-cristobalite converges more
quickly than that of $\alpha$-quartz. For $\alpha$-quartz, it is
difficult to define the convergence criterion to choose the cutoff
distance for the accurate lattice thermal conductivity calculation.

For a purpose of the rough estimation, any choice of the cutoff distance
and supercell size seems acceptable. In the case of $\alpha$-quartz and
$\alpha$-cristobalite, the first nearest neighbor distance is well
isolated because of SiO$_4$ tetrahedra. This may be the reason why the
atomic interaction range effective to determine lattice thermal
conductivity is found to be short.

\bibliography{SiO2-alpha-quartz}
\end{document}